\documentclass[twocolumn,numberedappendix]{emulateapj}
\usepackage{graphicx}
\usepackage{hyperref}
\usepackage{amssymb}
\usepackage{amsmath}
\usepackage{units}
\usepackage{fancyvrb}
\newcommand{\figwidth}{3.25in}

\newcommand{\comments}[1]{}

\def\app#1#2{%
  \mathrel{%
    \setbox0=\hbox{$#1\sim$}%
    \setbox2=\hbox{%
      \rlap{\hbox{$#1\propto$}}%
      \lower1.1\ht0\box0%
    }%
    \raise0.25\ht2\box2%
  }%
}

\DefineShortVerb{\|}

\begin{document}

\title{Acceleration of Relativistic Electrons by
  MHD Turbulence:  Implications for Non-thermal Emission from Black
  Hole  Accretion Disks} 
\author{Jacob W. Lynn\altaffilmark{1,2}, Eliot Quataert\altaffilmark{2}, Benjamin D. G. 
  Chandran\altaffilmark{3}, \& Ian J. Parrish\altaffilmark{4}
  } 

\altaffiltext{1}{Physics Department, University of California,
  Berkeley, CA 94720; jacob.lynn@berkeley.edu} 
\altaffiltext{2}{Astronomy Department and Theoretical Astrophysics 
  Center, University of California, Berkeley, CA 94720} 
\altaffiltext{3}{Space Science Center and Department of Physics,
  University of New Hampshire, Durham, NH 03824}
\altaffiltext{4} {Canadian Institute for Theoretical
    Astrophysics, 60 St. George Street, University of Toronto,
  Toronto, ON M$5$S $3$H$8$, Canada}

\begin{abstract}
We use analytic estimates and numerical simulations of test particles interacting with magnetohydrodynamic (MHD) turbulence to show that subsonic MHD turbulence produces efficient second-order Fermi acceleration of relativistic particles.    This acceleration is not well-described by standard quasi-linear theory but is a consequence of resonance broadening of wave-particle interactions in MHD turbulence.   We provide momentum diffusion coefficients that can be used for astrophysical and heliospheric applications and discuss the implications of our results for accretion flows onto black holes.   In particular, we show that particle acceleration by subsonic turbulence in radiatively inefficient accretion flows can produce a non-thermal tail in the electron distribution function that is likely important for modeling and interpreting the emission from low luminosity systems such as Sgr A* and M87.
\\
\end{abstract}

\keywords{plasmas -- heating --  acceleration of particles -- accretion, accretion disks}

\section{Introduction}
\label{sec:introduction}

In the limit of low-frequency magnetohydrodynamic (MHD) fluctuations,
charged relativistic particles 
are accelerated by mirror forces resulting from magnetic
compressions \citep{Achterberg1981},
\begin{equation}
  \label{eq:mirrorForce}
  \frac{d p_\parallel}{dt} = \frac{p_\perp v_\perp}{2B}
  \nabla_\parallel \mid\boldsymbol{B} \mid,
\end{equation}
where  $\parallel$ and $\perp$ denote directions relative to the local magnetic field. 
In MHD, magnetic compressions are caused by slow modes and fast
modes, with slow modes containing most of the compressive energy in
subsonic turbulence. Because slow modes
propagate approximately along the magnetic field in most regimes, a
pure linear resonance with relativistic particles requires
$\omega \simeq k_\parallel v_p = k_\parallel v_\parallel$, or
equivalently $v_p \simeq
c$, where $v_p$ is the parallel phase velocity of slow modes. Thus
linear theory predicts no acceleration of 
high-energy particles by MHD-scale slow modes, because the resonance condition
cannot be satisfied.   As a result, fast modes have traditionally been
believed to be the dominant source of relativistic particle
acceleration by MHD-scale turbulent fluctuations
\citep{Achterberg1981,Miller1996}.   However,  subsonic
turbulence does not contain significant fast mode energy (see, e.g., \citealt{Yao2011a, Howes2012} for empirical constraints on the fast and slow mode energy in the solar wind).   This appears
to significantly limit the efficiency of relativistic particle
acceleration by MHD turbulence in many astrophysical environments.

In strong MHD turbulence, the waves comprising MHD turbulence are not
long-lived but instead have a decay time comparable 
to their linear period.  In this case,
the linear resonance is not the appropriate condition for
wave-particle interaction. Instead, the 
resonance is nonlinearly broadened 
\citep{Bieber1994a,Gruzinov1999b,Shalchi2004,Shalchi2004a,Qin2006,Yan2008a,Lynn2012}. 
Resonance broadening allows waves to interact with
relativistic particles when $\omega_{\rm nl} \gtrsim k_\parallel
c$, where $\omega_{\rm nl}^{-1}$ is the non-linear correlation time of
the turbulence.   In this paper, we estimate the resulting
particle acceleration analytically (\S
\ref{sec:transportProperties}) and numerically using simulations of
relativistic test particles interacting with MHD turbulence (\S
\ref{sec:numericalMethods} \& \ref{sec:numericalResults}).   Our
results are potentially relevant to a wide range of astrophysical
plasmas; in \S \ref{sec:conclusions} we briefly assess the
implications of our results for non-thermal emission from accretion
disks around black holes. 

\section{Relativistic momentum diffusion by low-frequency MHD
  turbulence} 
\label{sec:transportProperties}

We first provide an order of magnitude estimate of the momentum
diffusion coefficient for relativistic particles interacting with magnetic field compressions associated with slow modes (the case of fast modes is considered separately). The diffusion coefficient for a particle with reduced momentum
$\overline{p} \equiv p/mc$ may be estimated as
$D_{\overline{p}, \boldsymbol{k}} \sim f^2 
\delta t / c^2$ where $f \sim \overline{p}
c^2 k_\parallel \delta B_\parallel(\boldsymbol{k}) / B_0$ is the force felt by a particle
interacting with a given spatial scale labeled by $\boldsymbol{k}$,
$\delta
t \sim  \omega_{\rm nl}^{-1}$ refers to the timescale over which
wave-particle interactions are correlated, and $\delta B_\parallel(\boldsymbol{k})$ is the rms fluctuation in magnetic compressions on scale $\boldsymbol{k}$.   For a given $k_\perp$,
the total acceleration will be determined by the 
average of $k_\parallel^2 \delta B_\parallel(\boldsymbol{k})^2$ over $k_\parallel$,
limited to those 
$k_\parallel$ that satisfy the broadened resonance condition 
$k_\parallel \lesssim \omega_{\rm nl} /c $. Provided that $\delta 
B_\parallel(\boldsymbol{k})^2$ does not scale too steeply with $k_\parallel$,
parallel wavenumbers near $k_\parallel \sim \omega_{\rm nl} /c$ will dominate, resulting in a 
diffusion coefficient at fixed $k_\perp$ of order
\begin{equation}
  \label{eq:approxDiffusion}
  D_{\overline{p}, k_\perp} \sim {\overline{p}^2}  \, \frac{v_A}{c} \, \frac{\omega_{\rm nl} \, \delta B_\parallel(\boldsymbol{k_\perp})^2.}{B_0^2}
\end{equation}
where we have used the fact that  most of the turbulent energy in anisotropic MHD turbulence has $k_\parallel \lesssim \omega_{\rm nl}/v_A$ and that only a fraction $v_A/c$ of the energy in magnetic compressions satisfies the conditions required for efficient particle acceleration, namely $k_\parallel \lesssim \omega_{\rm nl} /c$ {\bf (this follows formally from the magnetic field power spectrum in eq. \ref{eq:gsTurbulence} below, which implies a parallel energy spectrum of $dE/d \ln k_\parallel \propto k_\parallel$).}   Equation \ref{eq:approxDiffusion} shows that the scaling of
$\omega_{\rm nl} \delta B_\parallel(\boldsymbol{k_\perp})^2$ with $k_\perp$
determines which $k_\perp$ dominates. For the strong MHD power
spectrum of \citet{Goldreich1995}, $\omega_{\rm nl} \propto
k_\perp^{2/3}$ and $\delta B(\boldsymbol{k_\perp}) \propto
k_\perp^{-1/3}$, so that all scales contribute equally, provided that
they can satisfy $k_\parallel \lesssim \omega_{\rm nl} / c$ (which
favors long-wavelength fluctuations).   

More formally,  the
resonance-broadened diffusion coefficient for the reduced momentum is
given by \citep{Dupree1966,Weinstock1969}
\begin{equation}
  \label{eq:broadenedDiffusion}
  D_{\overline{p}_\parallel} = \frac{\overline{p}_\perp^2 v_\perp^2}{4 B_0^2} \int
  d^3k k_\parallel^2 I_B(\boldsymbol{k}) R(\boldsymbol{k}), 
\end{equation}
where $\overline{p}_\perp$ is the perpendicular component of the
dimensionless momentum, $I_B(\boldsymbol{k})$ is
the 3D power spectrum of magnetic field 
fluctuations, and $R(\boldsymbol{k})$ is a resonance function that
describes the time-averaged interaction of a test particle with waves
at a given $\boldsymbol{k}$.  Quantitatively, $R(\boldsymbol{k}) = \Re \int_0^\infty dt 
\exp{[\imath (\omega(\boldsymbol{k}) - v_\parallel k_\parallel) t]} \,
f(t)$, where $f(t)$ is the time correlation function for wave-particle
interactions. $R(\boldsymbol{k})$ is necessarily  
phenomenological, as it in principle depends on the momentum diffusion
coefficient itself.   Standard models in the literature assume that
waves decay as an exponential or Gaussian in time due to non-linear
interactions; e.g,  $f(t) = e^{-\omega_{\rm nl}^2 t^2}$ (a Gaussian
decay model), with a nonlinear decay frequency $\omega_{\rm nl}$.   We
focus on a Gaussian decay model favored by our previous test particle
simulations \citep{Lynn2012}.
These assumptions lead to \begin{equation} 
  \label{eq:gaussianResonance}
  R(\boldsymbol{k}) = \frac{\sqrt{\pi}}{2 \omega_{\rm nl}}
  \exp{\left[-\frac{k_\parallel^2 (v_\parallel-v_p)^2}{4 \omega_{\rm nl}^2}\right]}.
\end{equation}
where we have assumed for simplicity that the waves have a dispersion
relation $\omega = k_\parallel v_p$, a reasonable approximation for
anisotropic slow modes with $k_\perp \gg k_\parallel$.   Physically, equation \ref{eq:gaussianResonance} implies that for the particles to couple to the turbulent fluctuations, the frequency that the particles feel as they pass through a wave, $k_\parallel (v_\parallel - v_p)$, must be of order (or less than) the
nonlinear frequency.

To perform the calculation in Equation \ref{eq:broadenedDiffusion}, we
assume that the magnetic power spectrum of slow modes is given by
strong anisotropic turbulence,
\begin{equation}
  \label{eq:gsTurbulence}
  I(\boldsymbol{k}) \equiv \frac{\delta B_S^2 L^3}{12 \pi} (k_\perp L)^{-10/3}
  g\left(\frac{k_\parallel L^{1/3}}{k_\perp^{2/3}}\right),
\end{equation}
where $L$ is the outer scale of the cascade, $\delta B_S^2$ is the
total energy in slow mode magnetic fluctuations,  and $g(x) \simeq 1$ for $x \lesssim 1$ and falls off sharply to zero for $x \gtrsim 1$.    The cutoff $g(x)$ in equation \ref{eq:gsTurbulence}
represents the lack of power outside the
anisotropic Goldreich-Sridhar cone \citep{Goldreich1995} in weakly
compressible MHD turbulence,\footnote{{\bf \citet{Cho2002a} show using numerical simulations that $g(x) \propto \exp[-x]$.  Our results are insensitive to the precise functional form of g(x).}}  and the power spectrum normalization is
chosen so that 
$\int d^3k I(\boldsymbol{k}) \equiv \delta B_S^2 / 2$.
The nonlinear frequency in such turbulence is given by $\omega_{\rm nl} \simeq (v_A/L) (k_\perp 
L)^{2/3}$, which is of order the eddy turnover time on a given scale.

Finally, to simplify the resulting estimate, we  assume that the particles are 
relativistic ($\overline{p}_\perp \sim
\overline{p}_\parallel \gg 1$), and that wave speeds are
nonrelativistic.  To emphasize the net particle acceleration efficiency, we also present our results in terms of the total momentum diffusion coefficient, rather than the parallel momentum diffusion coefficient.   This implicitly assumes that modest pitch angle scattering isotropizes the distribution function (see \S \ref{sec:numericalResults}).  Under these approximations, the diffusion coefficient becomes 
\begin{align}
  \label{eq:slowModeIntermediateIntegral}
 &  D_{\overline{p}} = \frac{\sqrt{\pi} \overline{p}_\perp^2 v_\perp^2}{48 v_A}
   \frac{\delta B_S^2}{B_0^2}  \left(\frac{v_\parallel}{v}\right)^2  \times \\ & \int 
  dk_\parallel dk_\perp \frac{k_\parallel^2}{k_\perp^3} g\left(\frac{k_\parallel
  L^{1/3}}{k_\perp^{2/3}}\right)  \nonumber \exp{\left(-\frac{k_\parallel^2 L^2 v_\parallel^2}{4 (k_\perp L)^{4/3} v_A^2}\right)}.  
\end{align}
Both the exponential and
the $g(x)$ term in equation 
\ref{eq:slowModeIntermediateIntegral} 
have the effect of cutting off the interaction at 
high $k_\parallel$ for a given $k_\perp$.   When $v_A \ll v_\parallel$, the 
exponential cutoff will always be more constraining and requires
$k_\parallel \lesssim L^{-1} v_A/c (k_\perp L)^{2/3}$. Given this, we
find 
\begin{equation}
  \label{eq:slowModeDiffusion}
  {D_{\overline{p}}}= {\overline p^2} \frac{\pi}{24}
  \frac{v_A^2}{c L} \frac{\delta B_S^2}{B_0^2} 
  \frac{\left( 1 - \alpha^2 \right)^2}{\alpha^3}
  \ln{(k_{\rm max} L)},
\end{equation}
where we have rewritten the angular dependence in terms of $\alpha =
v_\parallel / v$, the particle pitch-angle 
cosine.   The result in equation \ref{eq:slowModeDiffusion} is similar to that derived earlier by \citet{Chandran2000}.   The restriction to parallel velocities much larger than $v_A$
corresponds to $\alpha \gg v_A / v \simeq v_A / c$.
Equation \ref{eq:slowModeDiffusion} corresponds to a particle acceleration time $t_{\rm
  a} \sim (\delta B_S/B_0)^{-2} (L/v_A) (c/v_A)$ .  Note  that this is independent of particle energy and is roughly the eddy turnover time divided by the fraction of the magnetic energy at low $k_\parallel$ that satisfies $k_\parallel \lesssim \omega_{\rm nl}/c$.

Equation \ref{eq:slowModeDiffusion} can be compared to the analogous result for acceleration of relativistic particles by fast modes.   The latter is predominantly via a linear resonance with highly oblique waves. Versions of this
calculation have been performed in many other contexts
\citep{Miller1996, Yan2002}, so we restrict ourselves to briefly summarizing the salient features here.

Relativistic test particles travelling along magnetic field lines can
experience a linear resonance with a highly 
oblique fast mode. The linear resonance function is a delta-function,
\begin{equation}
  \label{eq:linearFastModeResonance}
  R(\boldsymbol{k}) = \pi \delta{(k v_p \pm k_\parallel v_\parallel)},
\end{equation}
where $v_p$ is the isotropic phase velocity of fast modes
(approximately $v_A$ for $\beta \ll 1$ and $c_s$ for $\beta \gg 1$).
The power spectrum of fast modes in Alfv\'enic turbulence is 
believed to be isotropic. The spectral index is uncertain but 
the simulations of \citet{Kowal2010a} suggest a 1D power spectrum
$P(k) \sim k^{-2}$ (though possibly shallower; see \citealt{Cho2003, Chandran2005}).

Performing the integral in Equation \ref{eq:broadenedDiffusion} with
the fast mode linear resonance and isotropic power spectrum $\sim
\delta B_F^2 L^3 (k L)^{-\alpha}$ leads to \begin{equation}
  \label{eq:fastModeDiffusion}
  D_{\overline{p}} \sim \overline{p}^2 \frac{v_p^2}{c L} \frac{\delta B_F^2}{B_0^2}
  \int_{k_{\rm min}}^{k_{\rm max}} dk k^{1-\alpha},
\end{equation}
where $\delta B_F^2$ is the energy in magnetic compressions associated with fast modes.   For
$\alpha \sim 2$, the diffusion coefficient due to fast modes
(eq. \ref{eq:fastModeDiffusion}) is of the same form as that due to
slow modes (eq. \ref{eq:slowModeDiffusion}).  For $\beta \gg 1$, a comparison of these two expressions shows that ratio of the fast mode to slow mode diffusion coefficient is $\sim (c_s/v_A)^2 (\delta B_F/\delta B_S)^2$.  At high $\beta$, fast modes lose their magnetic compressibility (becoming simply sound waves), so that the magnetic energy $\delta B_F^2$ in fast modes decreases at fixed velocity amplitude, with $\delta B_F^2 \sim \rho \, \delta v_F^2/\beta$.   By contrast, $\delta B_S^2 \sim \rho \, \delta v_S^2$.   Thus the ratio of the fast to slow mode diffusion coefficients is in fact set by the relative turbulent energy in each mode.   For subsonic turbulence, slow modes will in general dominate the particle acceleration because there is significantly more energy in slow modes.   This depends, however, on the power spectrum of the fast modes.  If $\alpha \sim 3/2$ rather than $\alpha \sim 2$ (as in \citealt{Chandran2005}) then the fast mode acceleration efficiency can be greater than that of slow modes even given the overall lower energy density in fast modes.

\section{Numerical methods}
\label{sec:numericalMethods}

Our simulations consist of charged test particles evolving in
the macroscopic electric and magnetic fields of isothermal, subsonic
MHD turbulence. Apart from modifying the particle pusher for
relativistic test 
particles, which we describe below, our computational approach is 
identical to that of \cite{Lynn2012}. Dimensional quantities
throughout the paper are expressed in units of the sound speed $c_s$
and the box scale 
$L$, when not explicitly stated.

\subsection{Turbulence simulations}
\label{sec:turbSims}

We simulate ideal MHD turbulence with the Athena code
\citep{Stone2008}. We drive an incompressible turbulent velocity field
using an Ornstein-Uhlenbeck process, and allow 
compressible fluctuations to develop naturally. The OU process
has a characteristic autocorrelation time $t_{\rm OU}$. Fiducial
properties for the MHD simulations used in this work 
are summarized in Table \ref{table:fiducial}.   We show results from higher resolution calculations in Figure \ref{fig:Dp_vs_c} (discussed below) and find that the numerically determined diffusion coefficients are relatively independent of resolution. In our calculations,
the simulation box is extended along the
mean magnetic field, because otherwise the particles (which undergo
periodic boundary conditions) would interact with the same eddies
multiple times before the eddies decorrelate.

\begin{table}
  \begin{center}
    \caption{Summary of fiducial simulation properties}
    \begin{tabular}{ c c }
      \\
      \hline \hline
      Parameter & Value \\
      \hline \hline
      Resolution & $512\times128^2$ \\
      Volume ($L^3$) & $8 \times 2^2$ \\
      $\dot{\epsilon}$ ($c_s^3 / L$)\footnote{The turbulent energy input
        rate, corresponding to a sonic Mach number of $\simeq 0.35$.  Calculations with $\dot{\epsilon} = 0.01$ yield similar results.} & 0.1 \\
      $\beta$\footnote{Ratio of thermal to magnetic pressure.  Our calculations covered a range of $\beta \sim 0.1-10$.} & $1$ \\
      $t_{\rm OU} \, (L/c_s)\footnote{$t_{\rm OU}$
        refers to the correlation time in the Ornstein-Uhlenbeck turbulence forcing.}$  & $1.5$ \\
    $l_D$ ($L$)\footnote{Outer (driving) scale of the turbulence.} & $0.39$\\
     $\delta B_{\parallel}$ ($B_0$)\footnote{RMS fluctuation in the parallel magnetic field.} & $0.12$\\
     $N_{\mathrm{particles}}$ & $2^{11} \times 10^3 \simeq 2 \times 10^6$ \\
      $\Omega_0$ ($c_s/L$)\footnote{Test particle gyrofrequency.} & $2
      \times 10^5$ \\
               \end{tabular}
    \label{table:fiducial}
  \end{center}
\end{table}

\subsubsection{Measurement of turbulence properties}
\label{sec:turbulenceProperties}

An important property of our turbulence simulations for comparing test
particle results to analytical estimates is the rms deviation in
parallel magnetic field, $\delta B_{\parallel}$, since this sets the
magnitude of the magnetic mirror forces. We define this quantity as
the spatial rms average of $\mid \boldsymbol{B}\mid -
\mid\boldsymbol{B_0}\mid$, where 
$\boldsymbol{B_0}$ is the initial mean magnetic field. This is equivalent
to taking the local direction of the magnetic field as the
``parallel'' direction. The magnitude of $\delta B_\parallel$ depends
on the 
magnetic compressibility of the fast and slow modes at a given
$\beta$, in addition to their overall representation in the
turbulence. For our fiducial
simulation summarized in Table \ref{table:fiducial}, $\delta B_\parallel /B_0 \simeq 0.12$, while
simulations that have the same driving rate (and thus similar $v_{\rm 
  rms}$) but $\beta=0.3$ 
($\beta=3$) have $\delta B_\parallel /B_0 \simeq 0.05$ ($0.23$).  Note
that for higher $\beta$, the  energy in magnetic compressions is
larger. 

We also decompose the turbulent velocity field into the linear MHD
Alfv\'{e}n, slow, and fast modes, following the approximate Fourier
space method of \citet{Cho2003}.
For our fiducial simulation, 50\%, 45\%, and 5\% of the turbulent
energy is in the 
Alfv\'{e}n, slow, and fast modes respectively.  For both higher
and lower $\beta$, the proportion of energy in fast modes decreases
substantially (to less than $1\%$), while the slow mode energy remains
at the same order of magnitude. Thus it is broadly appropriate to
assume that all of the magnetic
field fluctuation energy is in the slow modes, and that
the particle acceleration is dominated by interactions with slow modes.  

\subsection{Test particle integration}
\label{sec:testParticles}

For a given fluid simulation, we simulate a statistical ensemble of
charged test particles which are initialized randomly throughout the
box of fully saturated turbulence. These test particles are evolved
according to the Lorentz force 
\begin{equation}
  \label{eq:lorentzForce}
  \frac{d \boldsymbol{\overline{p}}}{dt} = \frac{q}{m c} \boldsymbol{E} +
  \frac{q}{m c} \boldsymbol{\beta} \times \boldsymbol{B},
\end{equation}
where the dimensionless momentum $\boldsymbol{\overline{p}} \equiv
\boldsymbol{p} / mc$, and $\boldsymbol{\beta} = \boldsymbol{u}/c$ is
the particle's physical velocity. The $E$ 
and $B$-fields are those on the MHD grid, interpolated to the
particle's location using the triangular-shaped cloud
\citep{Hockney1981} method in space and time. For each simulation, we
also choose a numerical value for the speed of light $c$, which
affects the motion of the test particles. The choice of $c$ does not,
however, affect the turbulence.    Our
choice of  non-relativistic turbulence and relativistic test particles
is appropriate for studying high energy supra-thermal particle
acceleration. 

Particles are integrated
using the \citet{Vay2008} particle pusher, which is sympletic and
symmetric in time, and conserves energy and the magnetic moment
adiabatic invariant to machine precision in tests with constant
fields.\footnote{The \cite{Boris1970} pusher is not as accurate when fluid velocities are non-negligible fractions of the chosen value of $c$. In tests, these errors did not significantly affect our
  results, but we nevertheless prefer the Vay pusher for the relativistic
  case.}   We initialize the test particles with sufficiently high
gyrofrequencies that diffusion and heating is independent of
gyrofrequency; i.e. $\Omega \gg \omega$ where $\omega$ is the
frequency of any turbulent motions and $\Omega$ is the relativistic
gyrofrequency.  To calculate the momentum diffusion coefficients, we further initialize particles with a specific value of
$\overline{p}$, and generally take $\overline p_\perp = \overline
p_\parallel$. 
The momentum is defined with respect to
the bulk rest-frame of the simulation, though for the
relativistic particles we focus on, this choice is
unimportant because the particle velocities are much greater than the fluid velocities.

The  velocity diffusion
coefficients are calculated according to 
\begin{equation}
  \label{eq:diffusionDefinition}
  D_{\overline{p}} \equiv \frac{\langle \delta \overline{p}^2 \rangle} {2 \, \delta t},
\end{equation}
where the average is over many particles with the same initial
momentum. 
One subtlety is that because we initialize the particle momentum with
respect to the bulk rest frame of the simulation, they are not
initially moving with the local drift velocity. As they change their
motion to follow the drift velocity, the
particle momentum undergoes an initial transient ``jump'' which saturates 
at the rms velocity of the turbulence (times the Lorentz factor of the
test particle, for ultra-relativistic particles). We sidestep this
subtlety by fitting a linear function in $t$ to $\langle \delta
\overline{p}^2 \rangle$ at later times using
least-squares.

\section{Numerical results}
\label{sec:numericalResults}

In the analytic calculations summarized in \S \ref{sec:transportProperties}, the particles are assumed to diffuse primarily in $p_\parallel$, as expected for particles interacting with long wavelength, low frequency turbulent fluctuations.    In our numerical calculations, we find that particles undergo diffusion in both $p_\parallel$ and $p_\perp$ (or, equivalently, in total momentum $p$ and magnetic moment $\mu$).  This is true for both the relativistic calculations presented here and our earlier non-relativistic test particle calculations \citep{Lehe2009, Lynn2012}.   For the non-relativistic test particle calculations, the diffusion time for $\mu$ was somewhat longer than that for the total momentum $p$, while for the relativistic test particle results presented in this paper, the two timescales are comparable.    This diffusion in $\mu$ corresponds to an effective pitch angle scattering rate and may be due to violation of magnetic moment conservation by finite amplitude low frequency turbulent fluctuations \citep{Chandran2010}.   The theory for the latter has not been fully worked out for the $\beta \sim 1$ conditions we focus on here.   In what follows, we defer a detailed analysis of the diffusion in $\mu$ to future work and focus on the diffusion in total momentum $p$.

\begin{figure}
 \includegraphics[width=\figwidth]{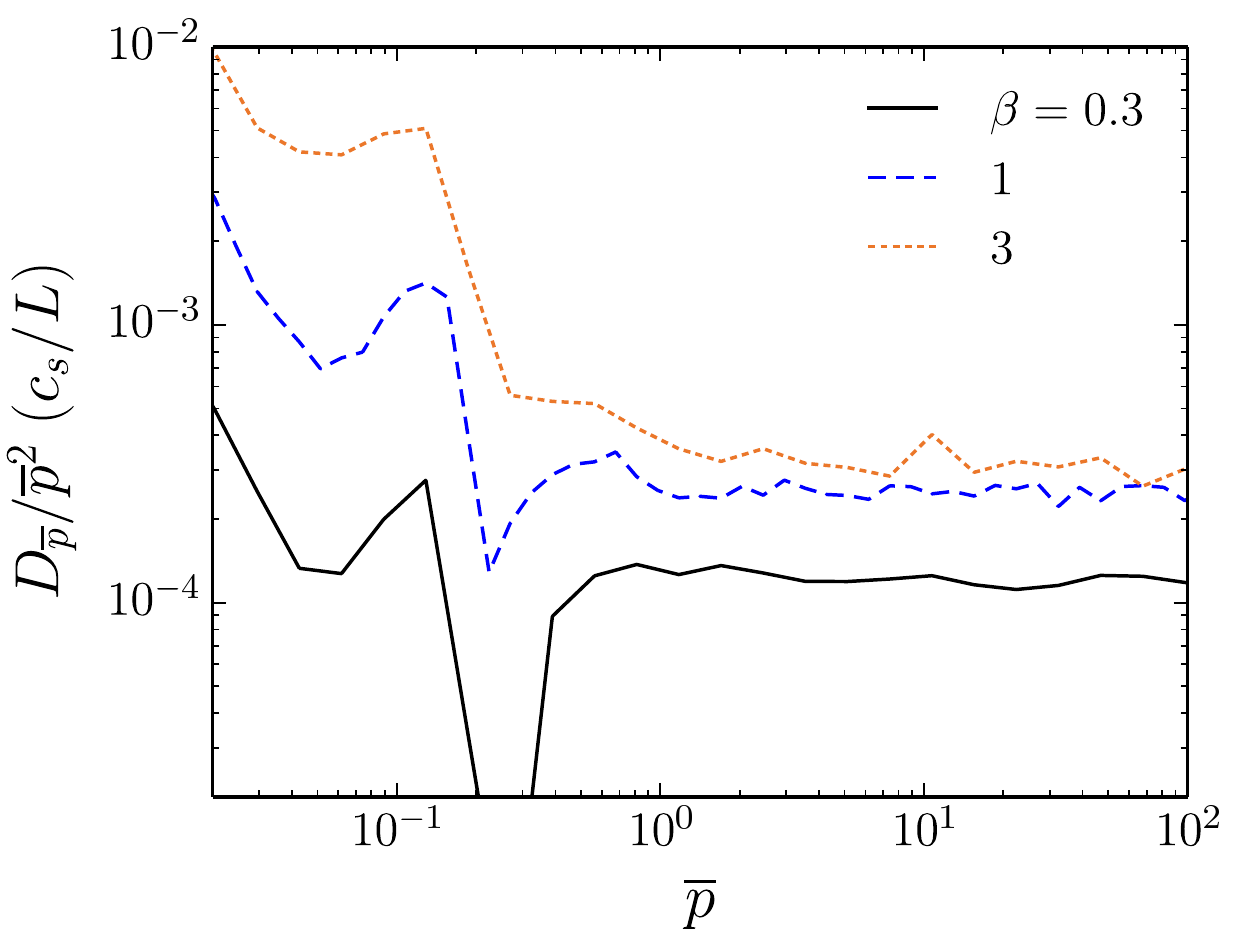}
  \caption{Test particle momentum diffusion coefficients as a function of $\overline p \equiv p/mc$ normalized by
    $\overline{p}^2$ for several values of $\beta$ (for fixed driving
    rate $\dot{\epsilon} = 0.01 c_s^3 / L$ and fixed $c=10 \,
    c_s$). For each $\overline p$,  the diffusion coefficient is measured for an ensemble of particles with ${\overline p}_{\perp} = {\overline
      p}_{\parallel} =  {\overline p}/\sqrt{2}$ that are initially at random positions in the turbulent box.   The numerical results demonstrate that $D_{\overline p} \propto    {\overline p}^2$ for    ultrarelativistic
    particles (${\overline p} \gg 1$), consistent with the analytic
    expectations from equation \ref{eq:slowModeDiffusion}.}
 \label{fig:Dp_vs_p}
  \
\end{figure}

Figure \ref{fig:Dp_vs_p} demonstrates that for relativistic test
particles, the momentum diffusion coefficient is robustly of the form
$D_{\overline p} \propto {\overline p}^2$. 
This is in contrast to the case of non-relativistic particles where
the diffusion coefficient for particles interacting with subsonic
turbulence is roughly $D \propto p$ for supra-thermal particles
\citep{Lynn2012}.  

The analytic results summarized in equation \ref{eq:slowModeDiffusion} also predict that for  relativistic particles, the magnitude of the diffusion coefficient depends on the magnetic compressibility of the turbulence and $v_A/c$.   We now test these expectations using our test particle simulations. 

\begin{figure}
 \includegraphics[width=\figwidth]{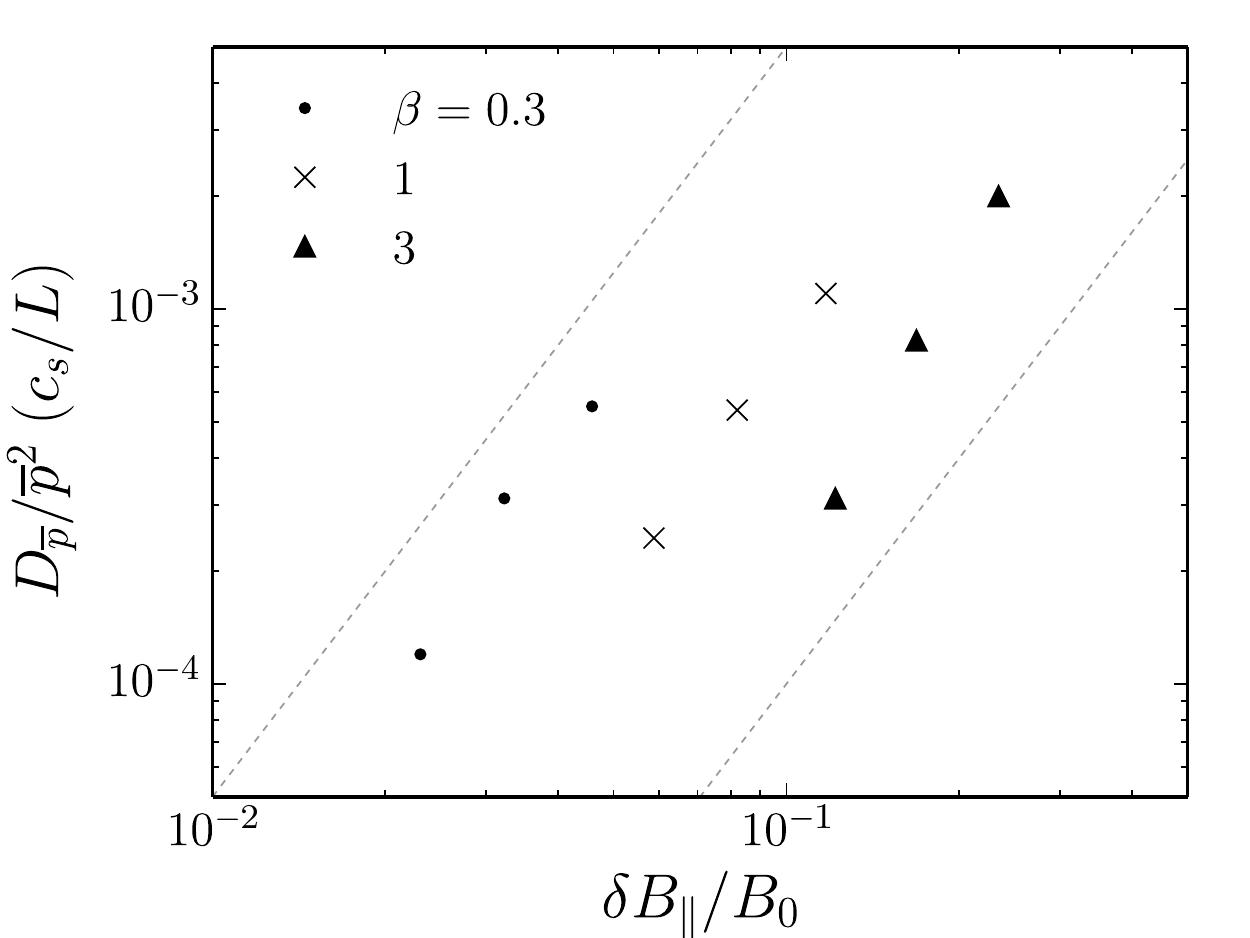}
  \caption{Test particle momentum diffusion coefficient for relativistic particles as a function of the strength of  the magnetic compressions $\propto \delta B_\parallel$, for several values of $\beta$.   The simulations have 
   $c = 10 \, c_s$ and different turbulent driving rates $\dot \epsilon$. The numerically determined
    diffusion coefficients are $\propto (\delta B_\parallel/B_0)^2$
    (light dashed lines, with arbitrary normalization), consistent with the  analytical predictions in \S \ref{sec:transportProperties}.}
  \label{fig:Dw_vs_dB2}
\end{figure}

Figure \ref{fig:Dw_vs_dB2} shows how the measured diffusion coefficient (normalized by $\overline{p}^2$) varies
with the rms magnetic field compression $\delta B_\parallel/B_0$. The latter is directly measured in the simulations as described in \S \ref{sec:turbulenceProperties}.    
Figure \ref{fig:Dw_vs_dB2} shows that the diffusion coefficient scales with the total turbulent energy in the magnetic field compressions, consistent with equation \ref{eq:slowModeDiffusion}.

\begin{figure}
 \includegraphics[width=\figwidth]{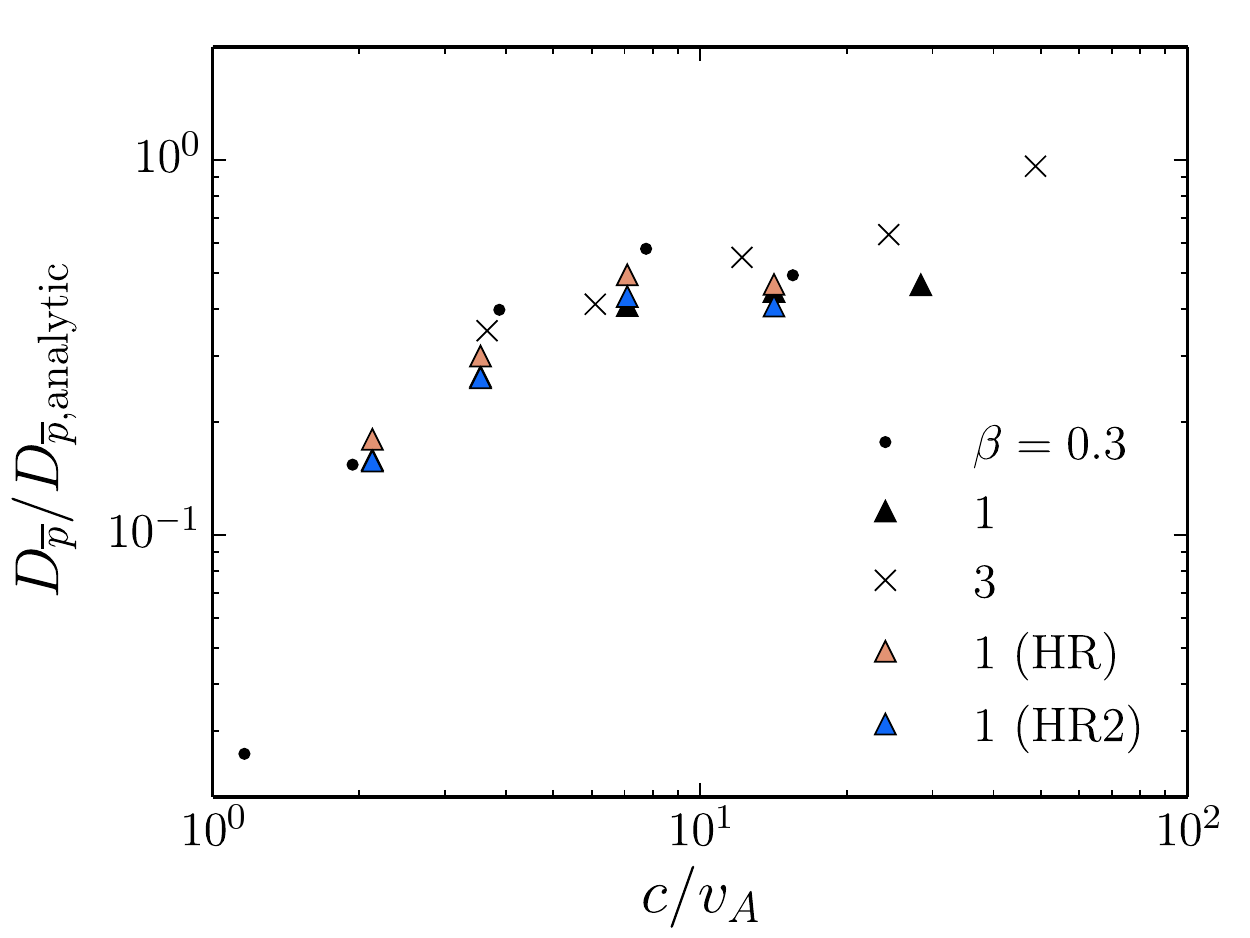}
\caption{Test particle momentum diffusion coefficients for relativistic particles normalized by
   the   analytic prediction of equation \ref{eq:slowModeDiffusion}
  for simulations with different $\beta$ and $v_A/c$.   For $c \gg
   v_A$, the results are well-described by the analytical predictions.
   The test particle calculations are for an ensemble of particles
   with ${\overline p}_{\perp} = {\overline p}_{\parallel} =
   {\overline p}/\sqrt{2}$.   The orange
   triangles (HR) are from a run with the fiducial number of grid
   cells but a larger turbulent driving scale, so that the
   turbulence has an inertial range that is 2 times larger.  The blue triangles (HR2) are for a higher resolution $1024\times256^2$ simulation.   Both resolution tests yield very similar results indicating that large scale turbulent fluctuations dominate the particle acceleration.} 
 \label{fig:Dp_vs_c}
\end{figure}

Figure \ref{fig:Dp_vs_c} shows the dependence of the measured diffusion
coefficient on the ratio $c/v_A$, for three different values of
$\beta$.  We reiterate that in each simulation, we
choose a value for the speed of light $c$ only for the purposes of evolving the test particles (the choice of $c$ has no impact on the properties of the turbulence).   The diffusion coefficients in Figure \ref{fig:Dp_vs_c} are normalized by the analytic
prediction in equation
\ref{eq:slowModeDiffusion}.\footnote{Specifically, we use
  $D_{\overline{p}} \simeq 0.4 \overline{p}^2 (\delta
  B_\parallel/B_0)^2 v_A^2/cL$ for the analytic prediction from
  equation \ref{eq:slowModeDiffusion}, with $\delta B_\parallel/B_0$
  calculated for each simulation, where we have used
$\ln \left( k_{\rm max} L \right) = \ln 64 \simeq 4.15$ for the fiducial simulation. For the
higher resolution simulations, the logarithmic factor is adjusted as
appropriate.}
For each $\beta$, $v_A$ and $c_s$
are fixed, so the x-axis in Figure \ref{fig:Dp_vs_c} corresponds to
different choices of $c$.    
For $c \gg v_A$, the results are reasonably consistent with the analytical
predictions.  In addition to the results for the fiducial simulation,
Figure \ref{fig:Dp_vs_c} also shows two other $\beta = 1$ simulations:
(1) one with the same resolution but a larger driving scale by a
factor of two, so that the inertial range is somewhat more extended
(HR)  (2) a second higher resolution 1024x256$^2$ simulation.    The
results for both of these other simulations are very similar to the
fiducial calculation, confirming that the large-scale fluctuations
that are well-resolved in a typical MHD simulation produce the
majority of the particle acceleration.     

\subsection{Long-time evolution of distribution function}
\label{sec:evolution}

\begin{figure*}
\includegraphics[width=7in]{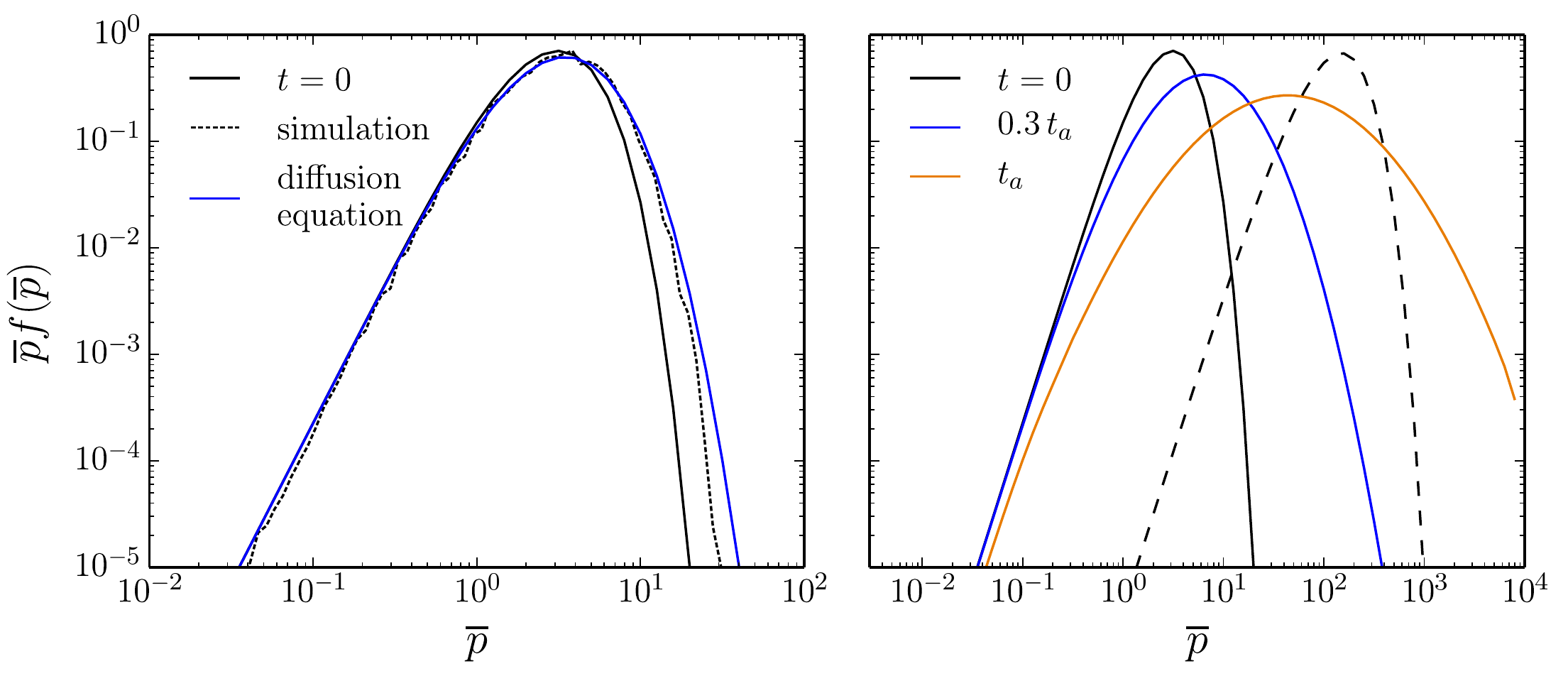}
 \caption{Evolution of the particle distribution function due to interaction with MHD turbulence.   \textit{Left panel:} Comparison of test particle simulations (dotted)
  with the numerical solution of the time dependent diffusion equation (solid blue) using
  a  momentum diffusion coefficient given by $D_{\overline p} \equiv
  \overline{p}^2/t_a$, where $t_a$ defines the acceleration time and is derived from the test particle results.   The particles are initially thermal and isotropic, and evolve over a
  long time baseline ($t = 20 \, L/c_s$ vs. $1 \, L/c_s$ for
  the calculations used to measure the diffusion coefficient). The numerical solution of the diffusion equation is shown at the same time and is in good agreement with the direct evolution of the test particles. \textit{Right panel:} Longer-time evolution of the numerical
  solution of the diffusion equation. The particles gain energy exponentially, with an
  e-folding time of approximately $0.25 \, t_a$.  On a comparable timescale, the distribution function develops a significant non-thermal tail.  For comparison, we also plot a thermal distribution with
  the same energy as the final distribution (black dashed curve), which highlights the substantial
  non-thermal tail at high energies.} 
  \vspace{0.15in}
  \label{fig:df_evol}
\end{figure*}

The diffusion coefficients shown in Figure \ref{fig:Dp_vs_c} correspond to particle acceleration times that are many eddy turnover times.   As a result, it is computationally intensive to directly simulate the long timescale evolution of the distribution function.   To study the latter, we instead separately solve the time dependent diffusion equation for the distribution function $f(p,t)$ using the momentum diffusion coefficients determined in our test particle calculations.   In particular, we begin with a Maxwellian distribution function
having $k_B T \sim m c^2$, i.e., $\langle \overline p \rangle \sim 1$ and evolve it subject to a diffusion coefficient given by $D_{\overline p} \equiv \overline p^2/t_a$ where $t_a$ defines the acceleration time.    Figure \ref{fig:df_evol} ({\em right panel}) shows the resulting distribution function at later times.  For comparison, we also show  a Maxwell-J{\"u}ttner distribution function (black dashed lines) that has the same total energy as the final distribution function in our diffusion calculations.     Figure \ref{fig:df_evol} shows that the distribution function quickly develops a significant non-thermal tail, on a timescale of $\sim 0.25 \, t_a$.   As a consistency check, the  {\em left panel} in Figure \ref{fig:df_evol} shows that over the timescale we can directly simulate the MHD turbulence with test particles, the evolution of the distribution function is indistinguishable from the solution of the momentum-space diffusion equation.

\
\section{Conclusions \& Implications}
\label{sec:conclusions}

Our results demonstrate that  subsonic MHD turbulence efficiently accelerates relativistic particles with a Fermi-like momentum diffusion coefficient $D_p \propto p^2$.  This is true for both $\beta \lesssim 1$ and $\beta \gtrsim 1$ and is thus a robust property of charged particles interacting with low frequency MHD turbulence.   We have restricted our analysis to particles whose (relativistic) cyclotron frequencies are  larger than the frequencies of the turbulent fluctuations.   In practice this limits our analysis to particles that are not too relativistic.   

Our key analytic result is that nonlinear broadening of quasi-linear resonances implies that slow modes in strong MHD turbulence can interact efficiently with relativistic particles, despite being
unable to satisfy the linear resonance condition (see \citealt{Chandran2000} for a similar result). 
{\bf In particular, resonance broadening allows long wavelength turbulent magnetic field compressions  satisfying $k_\parallel c \lesssim \omega_{\rm nl}$ to accelerate particles, where $\omega_{\rm nl}$ is the non-linear decay rate of the turbulence at a given scale.}

Because slow modes tend to be energetically more important than fast modes in subsonic turbulence, this suggests that interactions with slow modes may dominate the overall
particle acceleration by low-frequency, weakly compressible MHD
turbulence.  This is contrary to the standard quasi-linear theory
results in the literature (e.g., \citealt{Achterberg1981}).  However, the particle acceleration efficiency by fast  modes depends sensitively on their turbulent power spectrum, which is not fully understood.  In particular,  if the fast mode spectral index is $\alpha \sim 3/2$ (which is not the case in our simulations, though it is suggested by some studies), fast modes may be more efficient than slow modes at accelerating particles even if their total energy density is smaller (see eq. \ref{eq:fastModeDiffusion}). 

For relativistic particles, momentum diffusion of the form $D_p \propto p^2$ produces a power-law spectrum $dN/dp \propto p^{-1}$ so long as the acceleration time of particles (which is independent of particle energy) is shorter than the radiative loss timescale and the escape time from the acceleration region   \citep{Blandford1987}.     The total energy in the accelerated particle population depends on the efficiency with which `seed' relativistic particles are created.   Because suprathermal particle acceleration is  inefficient for non-relativistic particles interacting with MHD turbulence \citep{Lynn2012}, it is not clear if the net acceleration efficiency (by turbulent mechanisms alone) will be substantial for plasmas with non-relativistic temperatures, because the turbulence itself does not self-consistently seed relativistic particles.   By contrast, for relativistically hot plasmas, the formation of a non-thermal tail of relativistic particles by the mechanism studied here is likely to be quite efficient.    One particularly important application of our results is thus to accretion flows onto black holes, where the electrons can in some cases have $k T \gtrsim m_e c^2$ even though the disk turbulence itself is non-relativistic.     

\subsection{Implications for Black Hole Accretion Flows}

Weakly compressible MHD turbulence is
generic in black hole accretion flows as a consequence of the nonlinear
evolution and 
saturation of the magnetorotational instability \citep{Balbus1998}.
Non-thermal particle acceleration by such turbulence is of particular
astrophysical interest in at least two circumstances.  First, at low
accretion rates onto a black hole or neutron star, the accretion flow
can adopt a low-collisionality state in which much of the emission can
be dominated by a non-thermal population of electrons, if such a
population is present (e.g., \citealt{Yuan2003}).   Secondly, in luminous
radiatively efficient accretion flows, non-thermal emission from the
disk surface layers (a ``corona") can contribute significantly to the
synchrotron and high energy inverse Compton emission.  We briefly
discuss the implications of our results for these applications.    

The momentum diffusion coefficient calculated in \S
\ref{sec:transportProperties} and Figure \ref{fig:Dp_vs_c}
corresponds to a rate of energy gain given by \begin{equation} 
\label{eq:Edotacc}
\dot E_{\rm acc} \sim  \frac{c \, D_p}{p} \equiv A \, p \,
\frac{v_A^2}{L}\left(\frac{\delta B_\parallel}{ B_0}\right)^2 
\end{equation}
where $A$ is a dimensionless coefficient  that
encapsulates the efficiency of the particle acceleration and can be
calibrated using our test particle simulations.   In particular,
Figure \ref{fig:Dp_vs_c} corresponds to $A \sim 1/3$  for $c/v_A
\sim 10-100$, the values expected in the inner regions of accretion
disks around black holes. 
The exact value of $\delta B_\parallel/B_0$ in accretion disk turbulence is somewhat uncertain. For the $\beta \sim 10-100$ conditions expected, $\delta B_\parallel \sim 0.3 \, B_0$ is plausible.   However, the exact value depends in part on the effect of collisionless damping on the compressibility of accretion disk turbulence, which is not well understood.   Moreover, small-scale fluctuations generated by the mirror instability may contribute significantly to the magnetic field compressions in collisionless disks \citep{Kunz2014,Riquelme2014}.

The acceleration of particles by disk turbulence requires that the acceleration time is shorter than the viscous time.   Given the acceleration rate in equation \ref{eq:Edotacc} this is likely achieved in the inner regions close to the black hole.  In addition, the acceleration of particles by disk turbulence is limited by
radiative losses, in particular synchrotron and inverse Compton
emission.   Focusing on the former, 
we find that the maximum Lorentz factor of accelerated electrons
is given by 
\begin{equation}
\label{eq:gmax}
\gamma_{\rm max} \sim A \, \left(\frac{m_e}{m_p}\right) \tau_T^{-1} \left(\frac{\delta B_\parallel}{B_0}\right)^{2}
\end{equation} 
where $\tau_{T} \equiv \sigma_T n_e L$ is the Thompson optical depth
across the outer scale of the turbulent fluctuations $L$.   Equation \ref{eq:gmax} implies that non-thermal emission from accelerated
electrons is likely to be particularly important in low-luminosity
systems where $\tau_T \ll 1$.   As a concrete example, in models of
the emission from Sgr A*, $\tau_T \sim 10^{-5}-10^{-6}$ (e.g.,
\citealt{Yuan2003,Gammie2009}) so that $\gamma_{\rm max} \sim
100$.   This implies that the particle acceleration found here
may substantially modify the electron distribution function for
electrons that emit in the mm-infrared.    This is particularly
important to understand in the context of interpreting the variable
infrared emission and resolved mm images of Sgr A* (e.g.,
\citealt{Doeleman2008, Do2009}).      In the near future, more
detailed calculations of test particle electron acceleration in
shearing box simulations can be used to quantify the uncertain
dimensionless coefficient A in the above acceleration efficiency.

A second potential application of our results is to high energy emission from luminous accreting black holes, which can be produced by a combination of thermal and non-thermal processes.   However, phenomenological models of this emission suggest that $\tau_T \sim
0.1-1$ in the emission region \citep{Haardt1991,Esin1997}.   As a
result, it is unlikely that the particle acceleration found here is
sufficiently rapid to compete with radiative losses by synchrotron and
inverse Compton emission.    

\begin{acknowledgements}
This material is based on work supported by the National Science
Foundation Graduate Research Fellowship under Grant
No. DGE-1106400. This work was also supported in part by
NASA HTP grant NNX11AJ37G, NSF grant AST-1333682, a Simons Investigator award from the Simons Foundation, the David and Lucile Packard Foundation,
and the Thomas Alison Schneider Chair in Physics at UC Berkeley.
Computing time was provided by the National Science Foundation
TeraGrid/XSEDE resource on the Trestles and Kraken
supercomputer.
\end{acknowledgements}

\bibliographystyle{apj}
\bibliography{library,cites2}

\end{document}